\documentclass[conference]{IEEEtran}
\usepackage{graphicx}
\usepackage{amsmath}
\usepackage{amssymb}
\usepackage{amsthm, mathtools}
\usepackage{epstopdf}
\usepackage{cite}
\usepackage{xcolor}

\newtheorem{mydef}{Definition}
\newtheorem{mylmm}{Lemma}
\newtheorem{mythm}{Theorem}

\usepackage{algorithmicx}
\usepackage{algorithm}
\usepackage[noend]{algpseudocode}

\algnewcommand\algorithmicinput{\textbf{Input:}}
\algnewcommand\Input{\item[\algorithmicinput]}

\algnewcommand\algorithmicoutput{\textbf{Output:}}
\algnewcommand\Output{\item[\algorithmicoutput]}

\ifCLASSINFOpdf
\else
\fi

\begin{document}

\title{Generalizing Bottleneck Problems}
\author{
Hsiang Hsu$^*$, Shahab Asoodeh$^\dagger$, Salman Salamatian$^\ddagger$, and Flavio P. Calmon$^*$ \\
$^*$Harvard University, \{hsianghsu, fcalmon\}@g.harvard.edu, $^\dagger$University of Chicago, shahab@uchicago.edu, \\ $^\ddagger$Massachusetts Institute of Technology, salmansa@mit.edu
}

\maketitle

\begin{abstract}
Given a pair of random variables $(X,Y)\sim P_{XY}$ and two convex functions $f_1$ and $f_2$, we introduce two bottleneck functionals as the lower and upper boundaries of the two-dimensional convex set that consists of the pairs  $\left(I_{f_1}(W; X), I_{f_2}(W; Y)\right)$, where $I_f$ denotes $f$-information and $W$ varies over the set of all discrete random variables satisfying the Markov condition $W \to X \to Y$. Applying Witsenhausen and Wyner's approach, we provide an algorithm for computing boundaries of this set for $f_1$, $f_2$, and discrete $P_{XY}$. In the binary symmetric case, we fully characterize the set when (i) $f_1(t)=f_2(t)=t\log t$, (ii) $f_1(t)=f_2(t)=t^2-1$, and (iii) $f_1$ and $f_2$ are both $\ell^\beta$ norm function for $\beta \geq 2$. We then argue that upper and lower boundaries in (i) correspond to Mrs. Gerber's Lemma and its inverse (which we call Mr. Gerber's Lemma), in (ii) correspond to estimation-theoretic variants of Information Bottleneck and Privacy Funnel, and in (iii) correspond to Arimoto Information Bottleneck and Privacy Funnel.   
\end{abstract}

%
\IEEEpeerreviewmaketitle

\section{Introduction}
Few  information-theoretic  constructs  have  captured  the  attention of machine learning researchers and practitioners as  the  Information  Bottleneck  (IB) \cite{tishby2000information}. Given two correlated random variables $X$ and $Y$ with joint distribution $P_{XY}$, the goal of the IB is to determine a mapping $P_{W|X}$ that produces a new representation $W$ of $X$ such that 
(i) $W\to X\to Y$
and (ii) $I(W;Y)$ is maximized (information preserved) while minimizing $I(W;X)$ (compression). This tradeoff can be quantified by the Lagrangian functional
$\mathsf{B}(P_{XY}, \lambda) \triangleq \max\limits_{P_{W|X}} I(W;Y) - \lambda I(W;X)$.
The IB has proved useful in many machine learning problems, such as clustering \cite{tishby2001data} and natural language processing \cite{slonim2000document}. More recently, the IB framework has been used to analyze the training process of deep neural networks  \cite{tishby2015deep,shwartz2017opening}. 

In an inverse context, the Privacy Funnel (PF), introduced in \cite{calmon2017principal}, seeks to determine a mapping $P_{W|X}$ that minimizes $I(W;Y)$ (privacy leakage) while assuring $I(W; X)\geq x$ (revealing useful information). 
Analogously, the PF can be solved by considering the functional
$\mathsf{F}(P_{XY}, \lambda) \triangleq \min\limits_{P_{W|X}} I(W;Y) - \lambda I(W;X)$. 
The privacy funnel (and its variants) has shown to be useful in information-theoretic privacy \cite{calmon2017principal, calmon2015fundamental}.



The choice of  mutual information in both the IB and the PF frameworks does not seem to carry any specific ``operational" significance. It does, however, have a desirable practical consequence: it leads to self-consistent equations \cite[Eq.~28]{tishby2000information} that can be solved iteratively in the IB case. In fact, this property is unique to mutual information among many other information metrics \cite{harremoes2007information}.
Nevertheless, at least in theory, one can replace the mutual information with a broader family of measures based on $f$-divergences\footnote{Given two probability distributions $P\ll Q$ and a convex function $f: (0, \infty)\to \mathbb{R}$ with $f(1)=0$, $f$-divergences is $D_f(P \| Q) \triangleq \mathbb{E}_Q\left [ f(\frac{dP}{dQ}) \right ]$.}.

In this paper, we consider a wider class of \emph{bottleneck problems} which includes the IB and the PF. 
We define \emph{$f$-information} between two random variables $X$ and $Y$ as $I_f\left ( X; Y \right ) \triangleq D_f\left ( P_{XY} \| P_{X}P_{Y}\right )$, and introduce the following \emph{bottleneck functional}
\begin{equation}\label{def:fIB}
\begin{array}{r@{}l}
\mathsf{B}_{f_1, f_2}(P_{XY}, x)\triangleq \max\limits_{W \to X \to Y}  I_{f_2}\left ( W; Y \right )\mathsf{s.t.} I_{f_1}\left ( W; X \right ) \le x,
\end{array}
\end{equation}
and the \emph{funnel functional}
\begin{equation}\label{def:fPF}
\begin{array}{r@{}l}
\mathsf{F}_{f_1,f_2}(P_{XY}, x) \triangleq \min\limits_{W \to X \to Y}  I_{f_2}\left ( W; Y \right )\mathsf{s.t.} I_{f_1}\left ( W; X \right ) \ge x,
\end{array}
\end{equation}
where $f_1$ and $f_2$ are convex functions.
Different incarnations of $f$-information have already appeared, \textit{e.g.}, $T$-information in \cite{Polyanskiy} for $f(t)=|t-1|$.
These metrics possess ``operational" significance that are arguably more useful in statistical learning and privacy applications than mutual information.  For instance, total variation and Hellinger distance play important roles in hypothesis testing \cite{polyanskiy2014lecture} and $\chi^2$-divergence in estimation problems \cite{calmon2017principal}.
Formulations (\ref{def:fIB}) and (\ref{def:fPF}) for a broader class of divergences can be potentially useful to emerging applications of information theory in machine learning.


Computing (\ref{def:fIB}) and (\ref{def:fPF}) reduces to characterizing the upper and lower boundaries, respectively, of the two-dimensional set
\begin{equation}\label{eq:GBP}
    \Big\{\big(I_{f_1}(W; X), I_{f_2}(W;Y)\big):~W\to X\to Y \Big\}.
\end{equation}
It is worth mentioning that studying (\ref{eq:GBP}) is at the heart of the strong data processing inequalities \cite{calmon2015strong} as well as fundamental limits of privacy \cite{calmon2015fundamental}.
%
Witsenhausen \textit{et al.} \cite{witsenhausen1975conditional} investigated the lower boundary of a related set $\Big\{\left( H(X|W), H(Y|W)\right):~W \to X \to Y \Big\}$, where $H(\cdot)$ is the entropy function. In particular, they proposed an algorithm for analytically computing the lower boundary of $H(Y|W)$ based on a dual formulation. When $X$ is binary and $P_{Y|X}$ is a binary symmetric channel (BSC), the lower bound of the above set corresponds to the well-known Mrs.\ Gerber's Lemma \cite{el2011network}. Related convex techniques have also been used to characterize some network information theoretic regions \cite{nair2013upper}.

We generalize the approach in \cite{witsenhausen1975conditional} to study boundaries of \eqref{eq:GBP} for a broader class of $f$-information metrics, characterizing properties of new bottleneck problems of the form (\ref{def:fIB}) and (\ref{def:fPF}).
In particular, we investigate the estimation-theoretic variants of information bottleneck  and privacy funnel using $\chi^2$-divergence, which we call \emph{Estimation Bottleneck} and \emph{Estimation Privacy Funnel}, respectively. 
In the binary symmetric case, the upper boundary corresponds to the inverse of Mrs. Gerber's Lemma, which we call Mr. Gerber's Lemma. We further extend these lemmas for Arimoto conditional entropy \cite{sason2017arimoto}.



This paper is organized as follows. 
Section~\ref{sec:GP} introduces the geometry of bottleneck problems. In Section~\ref{sec:BP}, we formulate bottleneck problems and explore their use, and provide further applications on information inequalities in Section~\ref{sec:MGL}. 
%

\section{Geometric Properties}\label{sec:GP}
\subsection{Notation}
Let $X$ and $Y$ be two random variables having joint distribution $P_{XY}$ with supports $\mathcal{X} = \left [ m \right ] \triangleq \left \{ 1, \cdots, m \right \}$ and $\mathcal{Y} = \left [ n \right ]$, respectively. 
We denote by $P_X = \mathbf{q} \in \Delta_m$ the marginal probability vector with entries $\left [ P_X(1), \cdots, P_X(m) \right ]$, where $\Delta_m \triangleq \left \{ \mathbf{x} \in \mathbb{R}^m: \sum_{i=1}^m x_i = 1, x_i \ge 0 \right \}$ is a $m$-dimensional simplex. We denote by $\mathbf{T} \in \mathbb{R}^{n\times m}$ the stochastic matrix whose entries are the channel transformation $P_{Y|X}$, {\it i.e.} $\left [ \mathbf{T} \right ]_{i, j} = P_{Y|X}\left ( i| j \right )$; thus, $P_Y = \mathbf{T}\mathbf{q} \in \Delta_n$.
For a discrete random variable $W$ with support $\mathcal{W}$, let $\mathbf{p}_w = \left [ P_{X|W}(1|w), P_{X|W}(2|w), \dots, P_{X|W}(m|w) \right ]$, and let the marginal of $W$ be $P_W(w) = \alpha_w$. We denote by $h_m$ the entropy function, {\it i.e.} $h_m: \Delta_m \to \mathbb{R}$ with $h_m\left ( \mathbf{q} \right ) = - \sum_{i \in \left [ m \right ]} \mathbf{q}_i \log\mathbf{q}_i$ and $0\log0 \triangleq 0$. Finally, we denote the convex hull of a set $\mathcal{A}$ by $\mathsf{conv}\mathcal{A}$, and its boundary of a set by $\partial\mathcal{A}$. 

\subsection{Geometry of Bottleneck Problems}
Let $f: \Delta_m \to \mathbb{R}$ and $g: \Delta_n \to \mathbb{R}$ be continuous and bounded mappings over simplices of dimension $m$ and $n$, respectively. We study the set (\ref{eq:GBP}) by first considering a more general context, and then specialize it to different information metrics in following sections. We consider the tuple
\begin{equation}\label{def:expectation_pairs}
\Big( \mathbb{E}\big[ f\left ( \mathbf{p}_w \right ) \big], \mathbb{E}\big[ g\left ( \mathbf{T}\mathbf{p}_w \right ) \big] \Big),
\end{equation}
where $\mathbb{E}\left [ f\left ( \mathbf{p}_w \right ) \right ] = \sum_{w \in \mathcal{W}} \alpha_w f\left ( \mathbf{p}_w \right )$, and $\mathbb{E}\left [ g\left ( \mathbf{T}\mathbf{p}_w \right ) \right ] = \sum_{w \in \mathcal{W}} \alpha_w g\left ( \mathbf{T}\mathbf{p}_w \right )$. 
Recall that $X$, $Y$, and $W$ form the Markov chain $W\to X\to Y$. 
Therefore, we are interested in the following set for a fixed channel $\mathbf{T}$: 
\begin{eqnarray}
\mathcal{C}\left ( \mathbf{T} \right ) &\triangleq& \hspace{-0.5em} \{ \left ( \mathbf{q}, \mathbb{E}\left [ f\left ( \mathbf{p}_w \right ) \right ], \mathbb{E}\left [ g\left ( \mathbf{T}\mathbf{p}_w \right ) \right ] \right )| \nonumber\\
&&\hspace{-0.5em} \mathbf{p}_w \in \Delta_m, \sum_{w \in \mathcal{W}} \alpha_w \mathbf{p}_w = \mathbf{q}, \sum_{w \in \mathcal{W}} \alpha_w = 1\}.
\end{eqnarray}
Moreover, we define  $\mathcal{S}\left ( \mathbf{T} \right ) \triangleq \left \{ \left ( \mathbf{p}, f\left ( \mathbf{p} \right ), g \left ( \mathbf{T}\mathbf{p} \right ) \right )| \mathbf{p} \in \Delta_m \right \}$.
The next lemma is a direct generalization of \cite[Lemma 2.1]{witsenhausen1975conditional}.


\begin{mylmm}\label{lm:convex_compact}
$\mathcal{C}\left ( \mathbf{T} \right )$ is convex and compact with $\mathcal{C}\left ( \mathbf{T} \right ) = \mathsf{conv} \mathcal{S} \left ( \mathbf{T} \right )$.
In addition, all points in $\mathcal{C}\left ( \mathbf{T} \right )$ can be written as a convex combination of at most $m+1$ points of $\mathcal{S} \left ( \mathbf{T} \right )$; in other words, $\left | \mathcal{W} \right | \le m+1$.
\end{mylmm}

Let the upper and lower boundaries of $\mathcal{C}$ be denoted by $U_\mathbf{T}$ and $L_\mathbf{T}$, respectively, i.e., we have
\begin{eqnarray}
L_\mathbf{T}\left ( \mathbf{q}, x \right ) &\overset{\Delta}{=} & \inf \left \{ y | \left ( \mathbf{q}, x, y \right ) \in \mathcal{C}\left ( \mathbf{T}\right ) \right \}, \label{def:inf}\\ 
U_\mathbf{T}\left ( \mathbf{q}, x \right ) &\overset{\Delta}{=} & \sup \left \{ y| \left ( \mathbf{q}, x, y \right ) \in \mathcal{C}\left ( \mathbf{T}\right ) \right \}. \label{def:sup}
\end{eqnarray}
Under appropriate conditions on $x$ (depending on the choice of $f$),  $\left \{ y | \left ( \mathbf{q}, x, y \right ) \in \mathcal{C}\left ( \mathbf{T}\right ) \right \}$ is non-empty, and hence the compactness of $\mathcal{C}\left ( \mathbf{T}\right )$ allows one to replace infimum and supremum in (\ref{def:inf}) and (\ref{def:sup}) with minimum and maximum, respectively. Moreover, it follows from the convexity of $\mathcal{C}\left ( \mathbf{T}\right )$ that $L_\mathbf{T}\left ( \mathbf{q}, \cdot \right )$ is convex and $U_\mathbf{T}\left ( \mathbf{q}, \cdot \right )$ is concave. 

\subsection{Dual Formulations}\label{sec:GP_DA}
Since $\mathcal{C}(\mathbf{T})$ is a convex set, its upper and lower boundaries are equivalently represented by its supporting hyperplanes. We use the dual approach introduced in \cite{witsenhausen1975conditional} to evaluate  $L_\mathbf{T}\left ( \mathbf{q}, \cdot \right )$ and $U_\mathbf{T}\left ( \mathbf{q}, \cdot \right )$. 
For a given $\lambda$, define the conjugate function
\begin{equation}\label{def:dual}
L_\mathbf{T}^*\left ( \mathbf{q}, \lambda \right ) \triangleq \min\left \{ -\lambda x + y| \left (\mathbf{q},  x, y \right ) \in \mathcal{C}\left ( \mathbf{T}\right ) \right \}.
\end{equation}
Note that the graph of $L_\mathbf{T}(\mathbf{q}, \cdot)$ is the lower boundary of $\mathcal{C}\left ( \mathbf{T} \right )$. It then follows that the point $\left ( x, y \right )$ that achieves the minimum in (\ref{def:dual}) lies on the lower boundary of $\mathcal{C}\left ( \mathbf{T} \right )$ with supporting line of slope $\lambda$, and hence corresponds to a point $(x, L_\mathbf{T}\left ( \mathbf{p}, x \right ))$.

We now turn our attention to evaluating $L_\mathbf{T}^*\left ( \mathbf{q}, \cdot \right )$. Let 
\begin{equation}
\mathcal{S}_\lambda \left ( \mathbf{T}\right ) \overset{\Delta}{=} \left \{ \left ( \mathbf{p}, y - \lambda x \right )| \left ( \mathbf{p}, x, y \right ) \in \mathcal{S}\left ( \mathbf{T}\right ) \right \}.
\end{equation}
We observe that $\mathcal{S}_\lambda \left ( \mathbf{T}\right )$ is the graph of the function $\phi\left ( \cdot, \lambda \right )$ on $\Delta_m$ given by
\begin{equation}\label{def:phi}
\phi\left ( \mathbf{p}, \lambda \right ) = g\left ( \mathbf{T}\mathbf{p} \right ) - \lambda f\left ( \mathbf{p} \right ), \mathbf{p} \in \Delta_m.
\end{equation}
Since the mapping $y - \lambda x$ preserves convexity, we have
\begin{equation}
\mathcal{C}_\lambda \left ( \mathbf{T}\right ) \triangleq \mathsf{conv}\mathcal{S}_\lambda \left ( \mathbf{T}\right ) = \left \{ \left ( \mathbf{q}, y - \lambda x \right )| \left ( \mathbf{q}, x, y \right ) \in \mathcal{C}\left ( \mathbf{T}\right ) \right \}
\end{equation}
as the convex hull of the graph of $\phi\left ( \cdot, \lambda \right )$. 
Thus, $L_\mathbf{T}^*\left ( \mathbf{q}, \lambda \right )$ is given by the lower convex envelope of $\phi\left ( \cdot, \lambda \right )$ at $\mathbf{q}$.
The same would go, \textit{mutatis mutandis}, for $U_\mathbf{T}\left ( \mathbf{q}, x \right )$: 
its conjugate function $U_\mathbf{T}^*\left (\mathbf{q},\cdot \right )$, defined as 
$$U_\mathbf{T}^*\left (\mathbf{q},\lambda \right )\triangleq\max\left\{-\lambda x + y|\left (\mathbf{q}, x, y \right ) \in \mathcal{C}\left ( \mathbf{T}\right ) \right \}$$  
coincides with the upper concave envelope of $\phi\left (\cdot, \lambda \right )$ at $\mathbf{q}$.


These properties lead to a procedure for characterizing $L_\mathbf{T}\left (\mathbf{q},\cdot \right )$ and $U_\mathbf{T}\left (\mathbf{q},\cdot \right )$. We illustrate $L_\mathbf{T}\left (\mathbf{q},\cdot \right )$ in details and $U_\mathbf{T}\left (\mathbf{q},\cdot \right )$ will follow by using concave envelope instead.
For $L^*_\mathbf{T}\left ( \mathbf{q}, \lambda\right ) = z^*$, there are two scenarios:
\begin{enumerate}
    \item \textbf{Trivial case}: If $\left ( \mathbf{q}, z^* \right )$ is in both $\mathcal{S}_\lambda$ and $\mathcal{C}_\lambda$, then $z^* = g \left ( \mathbf{T} \mathbf{q} \right ) - \lambda f \left ( \mathbf{q} \right )$. In this case, $\left ( \mathbf{q}, x, L_\mathbf{T}\left ( \mathbf{q}, x \right ) \right )$ simply reduces to $\left (  \mathbf{q},  f \left ( \mathbf{q} \right ), g \left ( \mathbf{T} \mathbf{q} \right ) \right )$, and the optimal $W$ has $P_W\left ( w \right ) = 1$ for some $w$, independent of $X$.
    \item \textbf{Non-trivial case}: If $z^* \neq \phi \left ( \mathbf{q}, \lambda \right )$, then $\left ( \mathbf{q}, z^* \right ) \in \mathcal{C}_\lambda$ is the convex combination of points $\left ( \mathbf{p}_i, \phi \left ( \mathbf{p}_i, \lambda \right ) \right ) \in \mathcal{S}_\lambda$, with weights $\alpha_i$ where $i \in \left [ k \right ]$ for some $k \ge 2$, and $\sum_{i=1}^k \alpha_i = 1$. Then $\left ( \mathbf{q}, x, L_\mathbf{T}\left ( \mathbf{q},x \right ) \right )$ is given by $\sum_{i=1}^k \alpha_i \left ( \mathbf{p}_i, f\left ( \mathbf{p}_i \right ), g\left ( \mathbf{T}\mathbf{p}_i \right ) \right )$. Moreover, an optimal $W$ is attained by $P_W\left ( i \right ) = \alpha_i$ and $\mathbf{p}_w = \mathbf{p}_i$,  $i \in[k]$.
\end{enumerate}

Hence, the points on the graph of $L_\mathbf{T}(\mathbf{q}, \cdot)$ can be obtained by only considering the points of $\phi(\cdot, \lambda)$ which differs from its convex envelope $L^*_\mathbf{T}(\cdot, \lambda)$ since those are exactly the points where $W$ is not induced from the trivial case.
{\bf Algorithm \ref{algo:dual}} summarizes our previous discussions.

\begin{algorithm}[t!] 
\caption{Computing $\left ( x, L_\mathbf{T}\left ( \mathbf{q}, x \right ) \right )$ at slope $\lambda$}\label{algo:dual}
\begin{algorithmic}[1] 
\Input $\lambda, \mathbf{q}$
\Output $\left ( x, L_\mathbf{T}\left ( \mathbf{q}, x \right ) \right )$
\State Compute $\phi\left ( \mathbf{p}, \lambda \right )$, $\mathbf{p} \in \Delta_m$ 
\State $L_\mathbf{T}^*\left ( \mathbf{p}, \lambda \right )$ $\gets$ convex envelope of $\phi\left ( \mathbf{p}, \lambda \right )$
\If {$L_\mathbf{T}^*\left ( \mathbf{q}, \lambda \right ) = \phi\left ( \mathbf{q}, \lambda \right )$} 
\State \Return $\left ( f\left ( \mathbf{q} \right ), g\left ( \mathbf{T}\mathbf{q} \right ) \right )$
\Else
\State $\left ( \alpha_i, \mathbf{p}_i \right )_{i=1}^k \gets \left \{ \left ( a_i, \phi^{-1}\left ( b_i, \lambda \right ) \right )_{i=1}^k | b_i \in \phi\left ( \cdot, \lambda \right ), \right.$
\Statex \begin{flushright} $\left. \sum_{i=1}^k a_i b_i = L^*_\mathbf{T}\left ( \mathbf{q}, \lambda \right )\right \}$ \end{flushright}
\State \Return $\left ( \sum_{i=1}^k \alpha_i f\left ( \mathbf{p}_i \right ), \sum_{i=1}^k \alpha_i g\left ( \mathbf{T}\mathbf{p}_i \right )  \right )$
\EndIf
\end{algorithmic}
\end{algorithm}

\subsection{Matched Channels}
The geometry of bottleneck problems leads to intriguing properties of $\mathbf{p}_w$. 
Our previous discussions reveal that the points $\left \{ \mathbf{p}_i \right \}_{i=1}^k$ used to form the convex envelope of $\phi\left ( \cdot, \lambda \right )$ are special: they determine a channel $P_{X|W}$ such that for any distribution $P_W$, the resulting value of $\left ( \mathbf{q}, \mathbb{E}\left [ f\left ( \mathbf{p}_w \right ) \right ], \mathbb{E}\left [ g\left ( \mathbf{T} \mathbf{p}_w \right ) \right ] \right )$ is on the boundary of $\mathcal{C}\left ( \mathbf{T} \right )$ with supporting line of slope $\lambda$. In this case, we say that the points $\left \{ \mathbf{p}_i \right \}_{i=1}^k$ form a \emph{matched channel} for $\mathbf{T}$, $f$, and $g$. 
\begin{mydef}[Matched Channel]
For a fixed channel $P_{Y|X}$ and $f$, $g$, we say that $P_{X|W}$ is matched to $P_{Y|X}$ if there exists $P_W$ such that $\left | \mathcal{W} \right | \ge 2$ and 
\begin{equation}
\left ( \mathbb{E}\left [ f\left ( \mathbf{p}_w \right ) \right ], \mathbb{E}\left [ g\left ( \mathbf{T} \mathbf{p}_w \right ) \right ]  \right ) = \left ( x, L_\mathbf{T}\left ( \mathbf{q}, x \right ) \right ).
\end{equation}
\end{mydef}

Using an elementary result in convex geometry (see Lemma~\ref{lmm:2} in Appendix), we immediately have the following theorem. 
\begin{mythm} \label{thm:matched_channel}
Let $P_{X|W=w} = \mathbf{p}_w$ be a matched channel for $P_{Y|X}$. Then for any $P_W$, we have
\begin{equation}
\big ( \mathbb{E}\left [ f\left ( \mathbf{p}_w \right ) \right ], \mathbb{E}\left [ g\left ( \mathbf{T} \mathbf{p}_w \right ) \right ]  \big ) = \left ( x, L_\mathbf{T}\left ( \mathbf{q}, x \right ) \right ).
\end{equation}
\end{mythm}
\begin{proof}
See the Appendix.
\end{proof}

From Theorem~\ref{thm:matched_channel}, we know that for any distribution $P_X$, matched channels $P_{X|W}$ are entirely determined by the points on the curve $\phi\left ( \cdot, \lambda \right )$ whose convex combinations lead to the convex envelope of $\phi$ at $P_X$. It implies that as long as $\phi\left ( \cdot, \lambda \right )$ meets its convex envelope at $P_X$, small perturbation around $P_X$ does not change the matched channels $P_{X|W}$ but simply change the weight $\alpha_w$. Thus, optimal mappings $P_{W|X}$ are surprisingly robust to small errors in estimation of $P_X$, which could potentially give pragmatic advantages when applying bottleneck problems to real data. However, if $P_X$ changes, we can recover the matched channels by first solving $\alpha_i$ via $\mathbf{q}=\sum_{i=1}^{m+1} \alpha_i\mathbf{p}_w(i)$, where $\mathbf{p}_w(i) = P_{X|W=i}$, and then applying Bayes' rules.

Note that the properties above only hold when $f$ and $g$ do not depend on $P_X$. Specifically, matched channels do not exist for the cases studied in Section~\ref{sec:EB} and \ref{sec:EPF}.

\section{Generalizing Bottleneck Problems}\label{sec:BP}
In this section, we demonstrate how  the tools developed in Section~\ref{sec:GP} can be applied to new bottleneck problems of the form \eqref{def:fIB} and \eqref{def:fPF}. We then revisit the IB and PF, and also study their  estimation-theoretic variants. 

Consider the Markov chain $W \rightarrow X \rightarrow Y$. Our goal is to describe the achievable pairs of $f$-divergences
\begin{equation}\label{eq:BP_pairs}
\left ( D_{f_1}\left ( P_{WX}\|P_WP_X \right ), D_{f_2}\left ( P_{WY}\|P_WP_Y \right ) \right ).
\end{equation}
Observe that for a given $P_X$ we have
\begin{align}
&D_{f_1}\left ( P_{WX}\|P_WP_X \right ) \label{eq:conditional_Df}\\
&= \sum_{w \in \mathcal{W}} P_W(w) \left [  \sum_{x \in \mathcal{X}} P_X(x) {f_1}\left ( \frac{P_{X|W}(x|w)}{P_X(x)} \right )  \right ] \\
&= \sum_{w \in \mathcal{W}} P_W(w) D_{f_1} \left ( P_{X|W=w}\| P_X \right ),
\end{align}
 and hence $I_{f_1}(W; X)$ can be expressed as  
\begin{equation}
I_{f_1}(W; X) = \sum_{w \in \mathcal{W}} \alpha_w f\left ( \mathbf{p}_w \right ) = \mathbb{E}\big[ f\left ( \mathbf{p}_w \right ) \big],
\end{equation}
for some function $f$. Similarly, define $D_{f_2} \left ( P_{Y|W}\| P_Y \right ) = g(P_{Y|W})$, we have 
\begin{equation}
I_{f_2}(W;Y) = \sum_{w \in \mathcal{W}} \alpha_w g(\mathbf{T}\mathbf{p}_w) = \mathbb{E}\big[ g\left ( \mathbf{T}\mathbf{p}_w \right ) \big].
\end{equation}
Hence the corresponding set $\{(\mathbf{q}, I_{f_1}(W; X), I_{f_2}(W; Y))\}$ for varying $W$ has the same form as $\mathcal{C}(\mathbf{T})$. Letting 
\begin{eqnarray}
\phi\left ( \mathbf{p}, \lambda \right ) = D_{f_2}\left ( \mathbf{T}\mathbf{p}\|P_Y \right ) - \lambda D_{f_1}\left ( \mathbf{p}\|P_X \right ),
\end{eqnarray}
we can thus apply Algorithm~\ref{algo:dual} to characterize  $\mathsf{B}_{f_1, f_2}(P_{XY}, \cdot)$ and $\mathsf{F}_{f_1, f_2}(P_{XY}, \cdot)$.


Next, we show that how the usual IB and PF fit in our formulation, and study their estimation-theoretic counterparts. We note, however, that the previous analysis does not require $f_1=f_2$. 

\subsection{Information Bottleneck}\label{subsec:IB}
Assuming $f_1\left ( t \right ) = f_2\left ( t \right ) = t\log t$ in the bottleneck functional \eqref{def:fIB}, we have $D_{f_1}\left ( P_{WX}\|P_WP_X \right ) = I\left ( W; X \right )$ and $D_{f_2}\left ( P_{WY}\|~P_WP_Y \right ) = I\left ( W; Y \right )$. Thus, the set of points $\left \{ \left ( x, U_\mathbf{T}\left ( \mathbf{q}, x \right ) \right )| 0 \le x \le H\left ( X \right ) \right \}$
corresponds to the set of solutions of the IB problem. 

It is worth mentioning that the same geometric approach can also be applied directly to entropy functions which also leads to the IB formulation. In fact, this is exactly the setting studied in \cite{witsenhausen1975conditional}. Specifically, choosing $f_1 = h_m$ and $f_2 = h_n$, the set of points
\begin{equation}\label{eq:IB_2}
\left \{ \left ( h_m\left ( \mathbf{q} \right ) - x, h_n\left ( \mathbf{T}\mathbf{q} \right ) - L_\mathbf{T}\left ( \mathbf{q}, x \right ) \right )| 0 \le x \le h_m\left ( \mathbf{q} \right ) \right \}
\end{equation}
also corresponds to the set of solutions of the IB.
The IB is closely related to strong data processing inequalities. See \cite[Proposition 2]{calmon2015strong} for more details in the case of the BSC (see also Fig.~\ref{fig:BP_MGL} (right)).

\subsection{Privacy Funnel}\label{sec:PF}
Assuming $f_1\left ( t \right ) = f_2\left ( t \right ) = t\log t$ in the funnel functional \eqref{def:fPF}, the set 
$\left \{\left ( x, L_\mathbf{T}\left ( \mathbf{q}, x \right ) \right )| 0 \le x \le H\left ( X \right ) \right \}$ corresponds to the set of solutions 
of the privacy funnel, introduced in \cite{calmon2017principal}.
Equivalently, using the entropy function as in Section~\ref{subsec:IB}, the set of points 
\begin{equation}
\left \{ \left ( h_m\left ( \mathbf{q} \right ) - x, h_n\left ( \mathbf{T}\mathbf{q} \right ) - U_\mathbf{T}\left ( \mathbf{q}, x \right ) \right )| 0 \le x \le h_m\left ( \mathbf{q} \right ) \right \}
\end{equation}
also corresponds to the set of solutions, which follows from the fact that $U_\mathbf{T}\left ( \mathbf{q}, \cdot \right )$ is monotonically non-decreasing; see Fig.~\ref{fig:BP_MGL} (right).

\subsection{Estimation Bottleneck}\label{sec:EB}
One can move away from the usual IB and define new bottleneck problems by considering different functions $f_1$ and $f_2$.
For instance,  
if $f_1\left ( t \right ) = f_2\left ( t \right ) = t^2 - 1$, then the corresponding $f$-information, called $\chi^2$-information, is  defined as 
\begin{equation}
I_{f_1}(W; X)=\chi^2\left ( W; X \right ) \overset{\Delta}{=} \mathbb{E}\left [ \left ( \frac{P_{W, X}\left ( W, X \right )}{P_{W}\left ( W \right )P_{X}\left ( X \right )} \right ) \right ] - 1,
\end{equation}
We simplify the notation in (\ref{def:fIB}) for $\chi^2$-information as
\begin{eqnarray}\label{Eq:EB}
\mathsf{B}_{\chi^2}\left ( P_{XY}, x \right ) \triangleq \max\limits_{W \to X \to Y}  \chi^2\left ( W; Y \right )\mathsf{s.t.} \chi^2\left ( W; X \right ) \le x,
\end{eqnarray}

The reason to specifically study $\chi^2$- information are two-fold. First, it has been shown in \cite{calmon2017principal} that $\chi^2\left ( X; Y \right ) = \sum_{i=1}^d \lambda_i\left ( X; Y \right )$, where $\lambda_i\left ( X; Y \right )$ is the $i^\text{th}$ principal inertia component (PIC) of $P_{X, Y}$ and $d = \min\left \{ \left | \mathcal{X} \right |, \left | \mathcal{Y} \right | \right \} - 1$. Moreover, if the PICs between $X$ and $Y$ are large, then the minimum mean square error (MMSE) $\mathsf{mmse}\left ( X| Y \right )$ of estimating $X$ given $Y$ will be small \cite[Theorem 1]{calmon2017principal}, thereby making reliable estimations. 
Hence, if the goal of an estimation problem is to minimize $\mathsf{mmse}\left(Y|W \right)$, we can equivalently consider maximizing $\chi^2\left ( Y; W \right )$.
Second, following the spirit of the IB, we also add the constraint $\chi^2(W; X)\le x$ for the new representation $W$, as $\chi^2$-divergence serves as sharp bounds for any $f$-divergence \cite{makur2015bounds}.


Due to the above connection between $\mathsf{B}_{\chi^2}\left ( P_{XY}, x \right )$ and estimation problems, we call $\mathsf{B}_{\chi^2}\left ( P_{XY}, x \right )$ \emph{Estimation Bottleneck (EB)} problem. Clearly,  the set of points $\left \{ \left ( x, U_\mathbf{T}\left ( \mathbf{q}, x \right ) \right )| 0 \le x \le m-1 \right \}$ corresponds to the set of solutions of \eqref{Eq:EB}; see Fig.~\ref{fig:BP_MGL} (left). The bound $x \le m-1$ comes from the PIC analysis \cite{calmon2017principal}.

\subsection{Estimation Privacy Funnel}\label{sec:EPF}
\begin{figure}[t!]
	\centering
	\includegraphics[width=.5\textwidth]{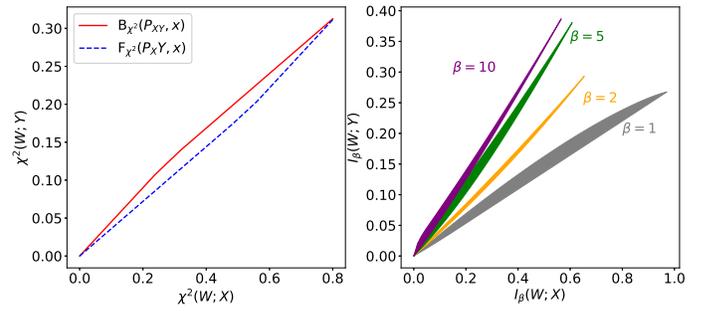}
	\caption{$\mathbf{T}$ follows BSC with crossover probability $\delta = 0.1$ and $P\{ X=1 \} = q = 0.1$. \textbf{Left}: The estimation bottleneck and privacy funnel. \textbf{Right}: Set of achievable pairs of Arimoto mutual information $\left \{  I_\beta(W;X), I_\beta(W;Y) \right \}$ for BSC with $\delta = 0.2$ and $q = 0.4$. Note that when $\beta=1$, the upper and lower boundaries correspond to the IB and the PF.}
	
	\label{fig:BP_MGL}
\end{figure}

Motivated by the connection between $\chi^2$-information and estimation problems mentioned in Section~\ref{sec:EB}, we propose
\begin{eqnarray}\label{def:chi_PF}
\mathsf{F}_{\chi^2}(P_{XY}, x) \triangleq \min\limits_{W \to X \to Y}  \chi^2\left ( W; Y \right )\mathsf{s.t.} \chi^2\left ( W; X \right ) \ge x,
\end{eqnarray}
where the privacy is measured in terms of MMSE. The practical significance of \eqref{def:chi_PF} is justified as follows. Suppose $Y$ represents private data (\textit{e.g.} political preferences) and $X$ (\textit{e.g.} movie rating) is correlated with $Y$. The main objective, formulated by (\ref{def:chi_PF}), is to construct a privacy-assuring mapping $P_{X|W}$ such that the information disclosed about $Y$ by $W$ is minimized, thus minimizing privacy leakage, while preserving the estimation efficiency that $W$ provides about $X$. Similarly, the solutions of the estimation privacy funnel \eqref{def:chi_PF} correspond to the set of points $\left \{ \left ( x, L_\mathbf{T}\left ( \mathbf{q}, x \right ) \right )| 0 \le x \le m-1 \right \}$ with $f_1\left ( t \right ) = f_2\left ( t \right ) = t^2 - 1$; see Fig.~\ref{fig:BP_MGL} (left). 


\section{Mrs. and Mr. Gerber's Lemmas}\label{sec:MGL}
The study of upper and lower boundaries of achievable mutual information pairs are essential in multi-user information theory \cite{el2011network}.
In the binary symmetric case, we not only rephrase Mrs. Gerber's lemma \cite{el2011network}, but also derive its counterpart for the PF. Furthermore, we discuss analogous results to Mrs. and Mr. Gerber's lemmas for Arimoto conditional entropy.

\subsection{Mr. Gerber's Lemma}
We apply the duality argument, developed in Section~\ref{sec:GP_DA}, for $L_\mathbf{T}$ and $U_\mathbf{T}$ to characterize the IB and the PF in the binary symmetric case. In particular, let $P_{Y|X}$ be the BSC with crossover probability $\delta$ and $q = \mathsf{Pr}\left ( X = 1 \right ) \le \frac{1}{2}$. For $a \in \left [ 0, 1 \right ]$, denote $\bar{a} = 1 - a$. We denote by $h_b\left ( q \right ) $ the binary entropy function $h_2\left ( \left [ q, \bar{q} \right ] \right )$.

Let $f_1 = f_2 = h_2$.  It was shown in \cite{witsenhausen1975conditional} that 
\begin{equation}\label{eq:MGL}
L_\mathbf{T}\left ( q, x \right ) = h_b\left ( \delta \star h_b^{-1}\left ( x \right ) \right ), \forall x \in \left [ 0, h_b\left ( q \right ) \right ],
\end{equation}
where $h_b^{-1}: [0, 1]\to [0, \frac{1}{2}]$ is the inverse function of $h_b\left ( \cdot \right )$ and $a \star b \triangleq \left ( 1-a \right )b + \left ( 1-b \right )a$, for $a, b \in \left [ 0, 1 \right ]$. Eq.~(\ref{eq:MGL}) is well-known as \textit{Mrs. Gerber's Lemma (MGL)}. In this case, the matched channel is also a BSC with crossover probability $h_b^{-1}\left ( x \right )$. Using the approach outlined in Section~\ref{sec:GP}, we derive a counterpart result for the upper boundary $U_\mathbf{T}$, and call it \textit{Mr. Gerber's Lemma}.

\begin{mythm}[Mr. Gerber's Lemma]\label{thm:MRGL}
For $0 \le q \le \frac{1}{2}$, we have
\begin{equation}\label{eq:MRGL}
U_\mathbf{T}\left ( q, x \right ) = \alpha h_b\left ( \delta \star \frac{q}{z} \right ) + \bar{\alpha}h_b\left ( \delta \right ),
\end{equation}
where $x = \alpha h_b\left ( \frac{q}{z} \right )$ and $z = \max\left ( \alpha, 2q \right )$, $\alpha \in \left [ 0, 1 \right ]$.
\end{mythm}
\begin{proof}
See the Appendix.
\end{proof}

In summary, in the binary symmetric case, the set of solutions for the IB follows from Mrs. Gerber's Lemma (\ref{eq:MGL}) and is given by $\left \{ \left ( h_b\left ( q \right )-x, h_b\left ( q \star \delta \right ) - L_\mathbf{T}\left ( q, x \right ) \right ) \right \}$, and the set of solutions for the PF follows from Mr. Gerber's Lemma \eqref{eq:MRGL} and is given by $\left \{ \left ( h_b\left ( q \right )-x, h_b\left ( q \star \delta \right ) - U_\mathbf{T}\left ( q, x \right ) \right ) \right \}$. The upper and lower boundaries of the achievable pairs $\left \{ I\left ( W; X \right ), I\left ( W; Y \right ): W\to X\to Y \right \}$ are therefore characterized by Mrs.\ and Mr.\ Gerber's Lemmas, respectively.

\subsection{Achievable Pairs of Arimoto Conditional Entropy}
Beside $\chi^2$-divergence and the entropy functions, one can choose $\ell^\beta$-norm $\left \| \cdot \right \|_\beta$ for $f$ and $g$ in (\ref{def:expectation_pairs}), which results in Arimoto's version of conditional R\'enyi entropy (Arimoto conditional entropy) \cite{sason2017arimoto} of order $\beta \geq 2$:
\begin{equation}\label{eq:ACE}
H_\beta\left ( X| W \right ) \triangleq \frac{\beta}{1-\beta} \log \sum_{w \in \mathcal{W}} \alpha_w \left \| \mathbf{p}_w \right \|_\beta.
\end{equation}
When $\beta = 1$, we define $H_1(X|W)=H(X|W)$. Hence, the set of achievable Arimoto conditional entropy pairs $\left ( H_\beta\left ( X| W \right ), H_\beta\left ( Y| W \right ) \right )$ can be obtained by the nonlinear mapping:
\begin{equation}
\left ( x, y \right ) \mapsto \left ( \frac{\beta}{1-\beta}\log x, \frac{\beta}{1-\beta} \log y \right ), \left ( x, y \right ) \in \mathcal{C}\left ( \mathbf{T} \right ).
\end{equation}
With (\ref{eq:ACE}) at hand, Arimoto mutual information \cite{sason2017arimoto} of order $\beta \geq 2$ can be  defined as $I_\beta \left ( X; W \right ) \triangleq H_\beta \left ( X \right ) - H_\beta \left ( X|W \right )$, where $H_\beta\left ( X \right )$ is the R\'enyi entropy of order $\beta$.
Arimoto conditional entropy has been proven useful in approximating the minimum error probability of Bayesian $M$-ary hypothesis testing \cite{sason2017arimoto}. 

\subsection{Arimoto's Mr. and Mrs. Gerber's Lemmas}
Due to the importance of Arimoto conditional entropy \cite{sason2017arimoto}, we study the extensions of Mr. and Mrs. Gerber's Lemmas for Arimoto conditional entropy, naming them Arimoto's Mr. and Mrs. Gerber's Lemmas respectively.

Let $K_\beta\left ( X| W \right ) = \exp\left \{ \frac{1-\beta}{\beta} H_\beta\left ( X| W \right )\right \}$, and also $K_\beta\left ( X \right ) = \exp\left \{ \frac{1-\beta}{\beta} H_\beta\left ( X \right )\right \}$. Since $ H_\beta\left ( X| W \right ) \le H_\beta\left ( X \right )$ for $\beta \geq 2$ and the mapping $x \mapsto \exp\left \{ \frac{1-\beta}{\beta} x\right \}$ is strictly decreasing, we have $K_\beta\left ( X| W \right ) \ge K_\beta\left ( X \right )$. Define $L_\mathbf{T}\left ( q, x \right )$ and $U_\mathbf{T}\left ( q, x \right )$ respectively as the minimum and maximum of $K_\beta\left ( Y| W \right )$ when $K_\beta\left ( X| W \right ) = x$ for $K_\beta\left ( X \right ) \le x \le 1$. For simplicity, denote $K_\beta\left ( q \right ) = K_\beta\left ( X \right )$ if $X \sim \text{Bernoulli}\left ( q \right )$. In this case, following section~\ref{sec:GP_DA}, we have $\phi\left ( p, \lambda \right ) = K_\beta\left ( \delta \star p \right ) - \lambda K_\beta\left ( p \right )$, which leads to the following theorem.

\begin{mythm}[Arimoto's Mrs. Gerber's Lemma]\label{thm:GMRGL}
For $0 \le q \le 1/2$ and $\beta \geq 2$, let $\mathcal{L}^{(\beta)} \overset{\Delta}{=} \left \{ \left ( x, L_\mathbf{T}\left ( q, x \right ) \right )| K_\beta\left ( q \right ) \le x \le 1 \right \}$. Then, we have
\begin{equation}
\mathcal{L}^{(\beta)} = \left \{ \left ( K_\beta\left ( p \right ), K_\beta\left ( p \star \delta \right ) \right )| 0 \le p \le q \right \}.
\end{equation}
In particular, $\frac{\beta}{1-\beta} \log y = \min\limits_{W \to X \to Y} H_\beta\left ( Y| W \right )$ s.t. $H_\beta\left ( X| W \right ) \ge \frac{\beta}{1-\beta} \log x$ for $\left ( x, y \right ) \in \mathcal{L}^{(\beta)}$.
\end{mythm}
\begin{proof}
See the Appendix.
\end{proof}

Analogous to this theorem, we also obtain the following generalization of Mr.\ Gerber's Lemma.

\begin{mythm}[Arimoto's Mr. Gerber's Lemma]\label{cor:GMSGL}
For $0 \le q \le 1/2$ and $\beta \geq 2$, let $\mathcal{U}^{(\beta)} \overset{\Delta}{=} \left \{ \left ( x, U_\mathbf{T}\left ( q, x \right ) \right )| K_\beta\left ( q \right ) \le x \le 1 \right \}$. Then, we have
\begin{equation}
\mathcal{U}^{(\beta)} = \left \{ \left ( \bar{\alpha}+\alpha K_\beta\left ( \frac{q}{z} \right ), \alpha K_\beta\left ( \frac{q}{z} \star \delta \right )+\bar{\alpha}K_\beta\left ( \delta \right ) \right ) \right \},
\end{equation}
where $\alpha \in \left [ 0, 1 \right ]$, and $z = \max\left \{ \alpha, 2q \right \}$.
In particular, we have
$\frac{\beta}{1-\beta} \log y = \max\limits_{W \to X \to Y} H_\beta\left ( Y| W \right )$ s.t. $H_\beta\left ( X| W \right ) \le \frac{\beta}{1-\beta} \log x$ for $\left ( x, y \right ) \in \mathcal{U}^{(\beta)}$.
\end{mythm}
Consequently, for $\beta \geq 2$, Arimoto's Mrs. and Mr. Gerber's Lemmas jointly characterize the achievable sets $\left \{  I_\beta(W;X), I_\beta(W;Y):W\to X\to Y \right \}$; see Fig.~\ref{fig:BP_MGL} (right). 


\section{Final Remarks}\label{sec:CC}
In this paper, we study the geometric structure behind bottleneck problems, and generalize the IB and PF to a broader class of $f$-divergences. In particular, we consider estimation-theoretic variants of the IB and PF. Moreover, we show how bottleneck problems can be used to calculate the counterpart of Mrs. Gerber's lemma (called Mr. Gerber's Lemma), and derive versions of Mrs. and Mr. Gerber's lemmas for Arimoto conditional entropy. These results can be potentially useful for new applications of  information theory in machine learning.

\appendix
\subsection{Lemma~\ref{lmm:2}}
\begin{mylmm}[\cite{eggleston1966convexity}]\label{lmm:2}
Let $\mathcal{A}$ be a connected and non-empty subset of $\mathbb{R}^n$, and $\mathcal{B} = \mathsf{conv} \mathcal{A}$. Assume there exists $\mathbf{x} \in \partial \mathcal{B}$ such that $x \notin \mathcal{A}$, then there exists $\left \{  \mathbf{x}_i\right \}_{i=1}^m$, where $m \le n$ with $\mathbf{x} = \sum_{i \in \left [ m \right ]} \alpha_i \mathbf{x}_i$, $\mathbf{x}_i \in \partial \mathcal{A}$, $0 < \alpha_i < 1$, and $\sum_i \alpha_i = 1$. Furthermore, $\mathsf{conv} \left \{  \mathbf{x}_i\right \}_{i=1}^m \subseteq \partial \mathcal{B}$.
\end{mylmm}

\subsection{Proof of Theorem~\ref{thm:matched_channel} (Matched Channel)}
Recall that $L_\mathbf{T}\left ( \mathbf{q}, \cdot \right )$ is determined parametrically in $\lambda$ by the points where $\phi\left ( \cdot, \lambda \right )$ does not match its convex envelope $L^*_\mathbf{T}\left ( \cdot, \lambda \right )$. Thus, the columns of the channel transformation matrix of a matched channel correspond to extreme points $P_{X|W}\left ( \cdot| i \right ) = \mathbf{p}_i$ where $\phi\left ( \cdot, \lambda \right )$ matches $L^*_\mathbf{T}\left ( \cdot, \lambda \right )$. However, there exists $\alpha_i$ where the convex combination $\sum_{i \in \left [ k \right ]} \alpha_i \mathbf{p}_i = \mathbf{q}$ corresponds to a point $\phi\left ( \mathbf{q}, \lambda \right ) \neq L^*_\mathbf{T}\left ( \mathbf{q}, \lambda \right )$. Using lemma \ref{lmm:2}, any non-trivial convex combination of $\mathbf{p}_i$ will result in a point $\mathbf{q}$ which is on the convex envelope of $\phi\left ( \cdot, \lambda \right )$ and determines a corresponding point on the curve $L_\mathbf{T}\left ( \mathbf{q}, \cdot \right )$.

\subsection{Proof of Theorem~\ref{thm:MRGL} (Mr. Gerber's Lemma)}
Take $f = g = h_2$, we have $\phi\left ( p, \lambda \right ) = h_b\left ( p \star \delta \right ) - \lambda h_b\left ( p \right )$. For $\lambda \ge \left ( 1-2\delta \right )^2$, $\phi\left ( \cdot, \lambda \right )$ is convex in $p$, and $U^*_\mathbf{T}\left ( q, \lambda \right ) = h_b\left ( \delta \right )$. For $0 \le \lambda < \left ( 1-2\delta \right )^2$, $\phi\left ( \cdot, \lambda \right )$ is concave in a region centered at $p = \frac{1}{2}$, where it reaches a local maximum. Consequently, if $\phi\left ( \frac{1}{2}, \lambda \right ) < h_b\left ( \delta \right )$, the upper convex envelope of $\mathcal{S}_\lambda$ is the linear combination of $\left ( 0, \phi\left ( 0, \lambda \right ) \right )$ and $\left ( 1, \phi\left ( 1, \lambda \right ) \right )$ and $U_\mathbf{T}\left ( q, x \right ) = h_b\left ( \delta \right )$. Assuming $p \le \frac{1}{2}$, if $\phi\left ( \frac{1}{2}, \lambda \right ) > h_b\left ( \delta \right )$, then there exists $p_\lambda \in \left [ 0, \frac{1}{2} \right ]$ such that for $p \le p_\lambda$, $\left ( p, U^*_\mathbf{T}\left ( q, \lambda \right ) \right ) \in \mathcal{C}_\lambda$ is a convex combination of $\left ( 0, \phi\left ( 0, \lambda \right ) \right )$ and $\left ( p_\lambda, \phi\left ( p_\lambda, \lambda \right ) \right )$. Finally, if $\phi\left ( \frac{1}{2}, \lambda \right ) = h_b\left ( \delta \right )$, then any point on the upper convex envelope of $\mathcal{S}_\lambda$ also lies in $conv\left \{ \phi\left ( 0, \lambda \right ), \phi\left ( \frac{1}{2}, \lambda \right ), \phi\left ( 1, \lambda \right ) \right \}$.

Hence, assuming $p \le \frac{1}{2}$, the distribution $P_{W, X}$ that achieves $U_\mathbf{T}\left ( q, x \right )$ will be of two cases:
\begin{enumerate}
	\item $Pr\left ( X = 1| W = 0 \right ) = 0$, $Pr\left ( X = 1| W = 1 \right ) = \frac{p}{\alpha}$ with $Pr\left ( W = 1 \right ) = \alpha$, $2p \le \alpha \le 1$.
	\item $W$ assuming values in $\left \{ 0, 1, 2 \right \}$ with $Pr\left ( X = 1| W = 0 \right ) = 0$, $Pr\left ( X = 1| W = 1 \right ) = 1$, $Pr\left ( X = 1| W = 2 \right ) = \frac{1}{2}$, with $Pr\left ( W = 0 \right ) = 1-p-\frac{\alpha}{2}$, $Pr\left ( W = 1 \right ) = p-\frac{\alpha}{2}$, $Pr\left ( W = 2 \right ) = \alpha$ and $0 \le \alpha \le 2p$.
\end{enumerate}
Rearranging (1) and (2), the result in (\ref{eq:MRGL}) follows.

\subsection{Proof of Theorem~\ref{thm:GMRGL} (Arimoto's Mr. Gerber's Lemma)}
Since $\phi\left ( \cdot, \lambda \right )$ is convex for $\lambda \le \left ( 1 - 2\delta \right )^2$. For $\lambda > \left ( 1 - 2\delta \right )^2$, $\phi''\left ( \cdot, \lambda \right )$ is negative on an interval $\left [ p_\lambda, \bar{p_\lambda} \right ]$, symmetric at $p = \frac{1}{2}$, and is positive elsewhere with local maximum at $p = \frac{1}{2}$. By symmetry, the lower convex envelope of the graph $\phi\left ( \cdot, \lambda \right )$ is obtained by replacing $p = p_\lambda$ if $p \in \left [ p_\lambda, \bar{p_\lambda} \right ]$. Therefore, for a given $q \le \frac{1}{2}$, if $p_\lambda \ge q$, then $\left ( q, L_\mathbf{T}\left ( q, \lambda \right ) \right )$ is a convex combination of $\left ( p_\lambda, \phi\left ( p_\lambda, \lambda \right ) \right )$ and $\left ( \bar{p_\lambda}, \phi\left ( \bar{p_\lambda}, \lambda \right ) \right )$. Hence, we have $L_\mathbf{T}\left ( q, x \right ) = K_\beta\left ( p \star \delta \right )$ for $x = K_\beta\left ( p \right )$ and $0 \le p \le q$.


\bibliographystyle{IEEEtran}
\bibliography{IEEEabrv,Reference}

\end{document}